\documentstyle[12pt]{article}

\catcode`\@=11
\long\def\@makefntext#1{ 
\protect\noindent \hbox to 3.2pt {\hskip-.9pt
$^{{\ninerm\@thefnmark}}$\hfil}#1\hfill} 

\def\thefootnote{\fnsymbol{footnote}}
 \def\@makefnmark{\hbox to 0pt{$^{\@thefnmark}$\hss}}  

\def\ps@myheadings{\let\@mkboth\@gobbletwo
\def\@oddhead{\hbox{} 
\rightmark\hfil\ninerm\thepage}
\def\@oddfoot{}\def\@evenhead{\ninerm\thepage\hfil 
\leftmark\hbox{}}\def\@evenfoot{}
\def\sectionmark##1{}\def\subsectionmark##1{}}

\begin{document}

\newcommand{\symbolfootnote}{\renewcommand{\thefootnote}
	{\fnsymbol{footnote}}}
\renewcommand{\thefootnote}{\fnsymbol{footnote}}
\newcommand{\alphfootnote}
	{\setcounter{footnote}{0}
	 \renewcommand{\thefootnote}{\sevenrm\alph{footnote}}}

\newcounter{sectionc}\newcounter{subsectionc}\newcounter{subsubsectionc}
\renewcommand{\section}[1] {\vspace{0.6cm}\addtocounter{sectionc}{1}
\setcounter{subsectionc}{0}\setcounter{subsubsectionc}{0}\noindent
	{\bf\thesectionc. #1}\par\vspace{0.4cm}}
\renewcommand{\subsection}[1] {\vspace{0.6cm}\addtocounter{subsectionc}{1}
	\setcounter{subsubsectionc}{0}\noindent
	{\it\thesectionc.\thesubsectionc. #1}\par\vspace{0.4cm}}
\renewcommand{\subsubsection}[1]
{\vspace{0.6cm}\addtocounter{subsubsectionc}{1}
	\noindent {\rm\thesectionc.\thesubsectionc.\thesubsubsectionc.
	#1}\par\vspace{0.4cm}}
\newcommand{\nonumsection}[1] {\vspace{0.6cm}\noindent{\bf #1}
	\par\vspace{0.4cm}}

\newcounter{appendixc}
\newcounter{subappendixc}[appendixc]
\newcounter{subsubappendixc}[subappendixc]
\renewcommand{\thesubappendixc}{\Alph{appendixc}.\arabic{subappendixc}}
\renewcommand{\thesubsubappendixc}
	{\Alph{appendixc}.\arabic{subappendixc}.\arabic{subsubappendixc}}

\renewcommand{\appendix}[1] {\vspace{0.6cm}
        \refstepcounter{appendixc}
        \setcounter{figure}{0}
        \setcounter{table}{0}
        \setcounter{equation}{0}
        \renewcommand{\thefigure}{\Alph{appendixc}.\arabic{figure}}
        \renewcommand{\thetable}{\Alph{appendixc}.\arabic{table}}
        \renewcommand{\theappendixc}{\Alph{appendixc}}
        \renewcommand{\theequation}{\Alph{appendixc}.\arabic{equation}}
        \noindent{\bf Appendix \theappendixc #1}\par\vspace{0.4cm}}
\newcommand{\subappendix}[1] {\vspace{0.6cm}
        \refstepcounter{subappendixc}
        \noindent{\bf Appendix \thesubappendixc. #1}\par\vspace{0.4cm}}
\newcommand{\subsubappendix}[1] {\vspace{0.6cm}
        \refstepcounter{subsubappendixc}
        \noindent{\it Appendix \thesubsubappendixc. #1}
	\par\vspace{0.4cm}}

\def\abstracts#1{{
	\centering{\begin{minipage}{30pc}\tenrm\baselineskip=12pt\noindent
	\centerline{\tenrm ABSTRACT}\vspace{0.3cm}
	\parindent=0pt #1
	\end{minipage} }\par}}

\newcommand{\bibit}{\it}
\newcommand{\bibbf}{\bf}
\renewenvironment{thebibliography}[1]
	{\begin{list}{\arabic{enumi}.}
	{\usecounter{enumi}\setlength{\parsep}{0pt}
\setlength{\leftmargin 1.25cm}{\rightmargin 0pt}
	 \setlength{\itemsep}{0pt} \settowidth
	{\labelwidth}{#1.}\sloppy}}{\end{list}}

\topsep=0in\parsep=0in\itemsep=0in
\parindent=1.5pc

\newcounter{itemlistc}
\newcounter{romanlistc}
\newcounter{alphlistc}
\newcounter{arabiclistc}
\newenvironment{itemlist}
    	{\setcounter{itemlistc}{0}
	 \begin{list}{$\bullet$}
	{\usecounter{itemlistc}
	 \setlength{\parsep}{0pt}
	 \setlength{\itemsep}{0pt}}}{\end{list}}

\newenvironment{romanlist}
	{\setcounter{romanlistc}{0}
	 \begin{list}{$($\roman{romanlistc}$)$}
	{\usecounter{romanlistc}
	 \setlength{\parsep}{0pt}
	 \setlength{\itemsep}{0pt}}}{\end{list}}

\newenvironment{alphlist}
	{\setcounter{alphlistc}{0}
	 \begin{list}{$($\alph{alphlistc}$)$}
	{\usecounter{alphlistc}
	 \setlength{\parsep}{0pt}
	 \setlength{\itemsep}{0pt}}}{\end{list}}

\newenvironment{arabiclist}
	{\setcounter{arabiclistc}{0}
	 \begin{list}{\arabic{arabiclistc}}
	{\usecounter{arabiclistc}
	 \setlength{\parsep}{0pt}
	 \setlength{\itemsep}{0pt}}}{\end{list}}

\newcommand{\fcaption}[1]{
        \refstepcounter{figure}
        \setbox\@tempboxa = \hbox{\tenrm Fig.~\thefigure. #1}
        \ifdim \wd\@tempboxa > 6in
           {\begin{center}
        \parbox{6in}{\tenrm\baselineskip=12pt Fig.~\thefigure. #1 }
            \end{center}}
        \else
             {\begin{center}
             {\tenrm Fig.~\thefigure. #1}
              \end{center}}
        \fi}

\newcommand{\tcaption}[1]{
        \refstepcounter{table}
        \setbox\@tempboxa = \hbox{\tenrm Table~\thetable. #1}
        \ifdim \wd\@tempboxa > 6in
           {\begin{center}
        \parbox{6in}{\tenrm\baselineskip=12pt Table~\thetable. #1 }
            \end{center}}
        \else
             {\begin{center}
             {\tenrm Table~\thetable. #1}
              \end{center}}
        \fi}

\def\@citex[#1]#2{\if@filesw\immediate\write\@auxout
	{\string\citation{#2}}\fi
\def\@citea{}\@cite{\@for\@citeb:=#2\do
	{\@citea\def\@citea{,}\@ifundefined
	{b@\@citeb}{{\bf ?}\@warning
	{Citation `\@citeb' on page \thepage \space undefined}}
	{\csname b@\@citeb\endcsname}}}{#1}}

\newif\if@cghi
\def\cite{\@cghitrue\@ifnextchar [{\@tempswatrue
	\@citex}{\@tempswafalse\@citex[]}}
\def\citelow{\@cghifalse\@ifnextchar [{\@tempswatrue
	\@citex}{\@tempswafalse\@citex[]}}
\def\@cite#1#2{{$\null^{#1}$\if@tempswa\typeout
	{IJCGA warning: optional citation argument
	ignored: `#2'} \fi}}
\newcommand{\citeup}{\cite}

\def\fnm#1{$^{\mbox{\scriptsize #1}}$}
\def\fnt#1#2{\footnotetext{\kern-.3em
	{$^{\mbox{\sevenrm #1}}$}{#2}}}

\font\twelvebf=cmbx10 scaled\magstep 1
\font\twelverm=cmr10 scaled\magstep 1
\font\twelveit=cmti10 scaled\magstep 1
\font\elevenbfit=cmbxti10 scaled\magstephalf
\font\elevenbf=cmbx10 scaled\magstephalf
\font\elevenrm=cmr10 scaled\magstephalf
\font\elevenit=cmti10 scaled\magstephalf
\font\bfit=cmbxti10
\font\tenbf=cmbx10
\font\tenrm=cmr10
\font\tenit=cmti10
\font\ninebf=cmbx9
\font\ninerm=cmr9
\font\nineit=cmti9
\font\eightbf=cmbx8
\font\eightrm=cmr8
\font\eightit=cmti8


\newcommand{\auth}[1]{{ #1}}
\newcommand{\tsetit}[1]{{\it #1}}
\newcommand{\journal}[1]{{\it #1}}
\newcommand{\vol}[1]{{\bf #1}}

\newcommand{\tnote}[1]{}

\newcommand{\tseno}[1]{}

\begin{flushright}
Imperial/TP/93-94/4 \\
hep-ph/9310339 \\
19th October, 1993 \\
\end{flushright}

\centerline{\tenbf A NEW TIME CONTOUR FOR }
\baselineskip=16pt
\centerline{\tenbf THERMAL FIELD THEORIES\footnote{Talk
given at the ``3rd Workshop on Thermal Field Theories and their
applications'', August 15th - 27th, 1993, Banff, Canada.  To appear in
the proceedings, to be published by World Scientific}
}
\vspace{0.8cm}
\centerline{\tenrm T.S. EVANS\footnote{E-mail: T.Evans@IC.AC.UK}}
\baselineskip=13pt
\centerline{\tenit Blackett Lab., Imperial College, Prince Consort Road}
\baselineskip=12pt
\centerline{\tenit London, SW7 2BZ, U.K.}
\vspace{0.9cm}
\abstracts{A simpler new time contour for  path-ordered approaches to
real-time thermal field theories is presented.  In doing so the use
of a so called `asymptotic' condition as seen in existing derivations
is seen to be incorrectly applied but luckily unnecessary.
}

\vfil
\twelverm   
\baselineskip=14pt


In this talk, I shall be concerned with path ordered approaches to
equilibrium thermal field theory.  These include both PORTF
(path-ordered real-time formalisms) of the various
types\cite{TSEnrtf,NS,ctp,Raybook,LvW}
and the ITF (imaginary-time formalism
or Matsubara method)\cite{Raybook,LvW,ITF}.    The path-ordered
approach to a real-time formalism is to be distinguished from the Thermo
Field Dynamics type approaches to real-time thermal field theories.  The
latter includes the approach due to Umezawa and
collaborators\cite{Raybook,TFD} as well as the axiomatic field theory version
involving $C^\ast$-algebra methods\cite{LvW}, since the two approaches
are identical\cite{Oj}.  Thermo Field Dynamics and PORTF are related
but they have important differences\cite{LvW}.


The starting point for path-ordered approaches to thermal field theory
is the thermal generating functional, $Z[j]$,
\begin{equation}
Z[j] := {\rm Tr} \{ e^{-\beta H}
\exp \{ T_C \int_C d\tau \int d^3\vec{x} \; \; j(\tau,\vec{x})
\phi(\tau,\vec{x}) \}
\label{eZjdef}
\end{equation}
The sources $j$ are  coupled to the fields, here generically denoted
by $\phi$.  In principle there is a source for every field but for
simplicity this will be represented by a single $j\phi$ term.  These
sources are unphysical and are set to zero at the end of the
calculation.  The $T_c$ indicates that the fields are path ordered with
respect to the relative order of their time arguments along
a directed path, $C$, in the complex time
plane\cite{NS,LvW}.  Then, by using one's favourite method, such as
the path-integral\cite{NS,Raybook,LvW} or operator methods\cite{ctp},
one can obtain Feynman rules, the effective action or
whatever else is required.

In order for path-ordered methods to work, the path $C$  starts at
some arbitrary time, say $\tau_{in}$, and then must end at a time
$\tau_{out}=\tau_{in}-i\beta$.  It has been suggested on formal
grounds  that the path must also always have a decreasing imaginary
part but this limitation is not seen in the Feynman
rules.\footnote{For instance the  general curve I described
elsewhere\cite{TSEnrtf} can slope upwards yet the Feynman rules are
independent of this factor.}  Here we stay also within this
limitation.   Any $C$ satisfying these conditions may be choosen.
Physical  results are therefore independent of $C$.   One of the
great advantages of the path-ordered approach to thermal field theory
is that the different FTFT formalisms simply correspond to different
choices for $C$.

The Green functions which are generated from $Z[j]$ are path ordered
expectation values of fields where the fields are
ordered according to the position of their time arguments on $C$. Thus
the physics is encoded in different ways for different choices of $C$
and thus for different FTFT formalisms\cite{TSErtgf}.  All the
thermodynamic information can be obtained from calculating the
partition function $Z=Z[j=0]$.

\begin{figure}[htb]
\setlength{\unitlength}{0.0125in}%
\begin{picture}(490,360)(40,400)
\thicklines
\put( 80,480){\circle*{10}}
\put( 80,670){\circle*{10}}
\put(480,670){\circle*{10}}
\put(480,480){\circle*{10}}
\put( 80,420){\circle*{10}}
\put( 40,680){\vector( 1, 0){480}}
\put( 80,670){\vector( 1, 0){280}}
\put(480,670){\vector( 0,-1){ 70}}
\put(480,480){\vector(-1, 0){280}}
\put( 80,480){\vector( 0,-1){ 40}}
\put(360,670){\line( 1, 0){120}}
\put(480,600){\line( 0,-1){120}}
\put(200,480){\line(-1, 0){120}}
\put( 80,440){\line( 0,-1){ 20}}
\put(480,680){\line( 0, 1){ 15}}
\put( 80,680){\line( 0, 1){ 15}}
\put(280,420){\line( 1, 0){ 15}}
\put(280,400){\vector( 0, 1){360}}
\put(345,645){\makebox(0,0)[lb]{\Large $C_1$}}
\put( 50,450){\makebox(0,0)[lb]{\Large $C_4$}}
\put(285,685){\makebox(0,0)[lb]{\Large $0$}}
\put(285,490){\makebox(0,0)[lb]{\Large $-\imath(1-\alpha)\beta$}}
\put( 60,705){\makebox(0,0)[lb]{\Large $-{\cal T}/2$}}
\put(460,705){\makebox(0,0)[lb]{\Large ${\cal T}/2$}}
\put(305,410){\makebox(0,0)[lb]{\Large $ - \imath \beta$}}
\put(205,495){\makebox(0,0)[lb]{\Large $C_2$}}
\put(445,600){\makebox(0,0)[lb]{\Large $C_3$}}
\put(490,650){\makebox(0,0)[lb]{\Large $\Re e (\tau)$}}
\put(295,730){\makebox(0,0)[lb]{\Large $\Im m (\tau)$}}
\multiput(285,485)(-0.40000,-0.40000){26}{\makebox(0.4444,0.6667){\sevrm .}}
\multiput(275,485)(0.40000,-0.40000){26}{\makebox(0.4444,0.6667){\sevrm .}}
\put(240,420){\circle{10}}
\put(240,670){\circle{10}}
\put(245,560){\makebox(0,0)[lb]{\Large $C_{I}$}}
\put(240,560){\line( 0,-1){140}}
\put(240,670){\vector( 0,-1){110}}
\end{picture}
\caption{The curves for imaginary-time and old real-time approaches.}
\label{fitfortf}
\end{figure}
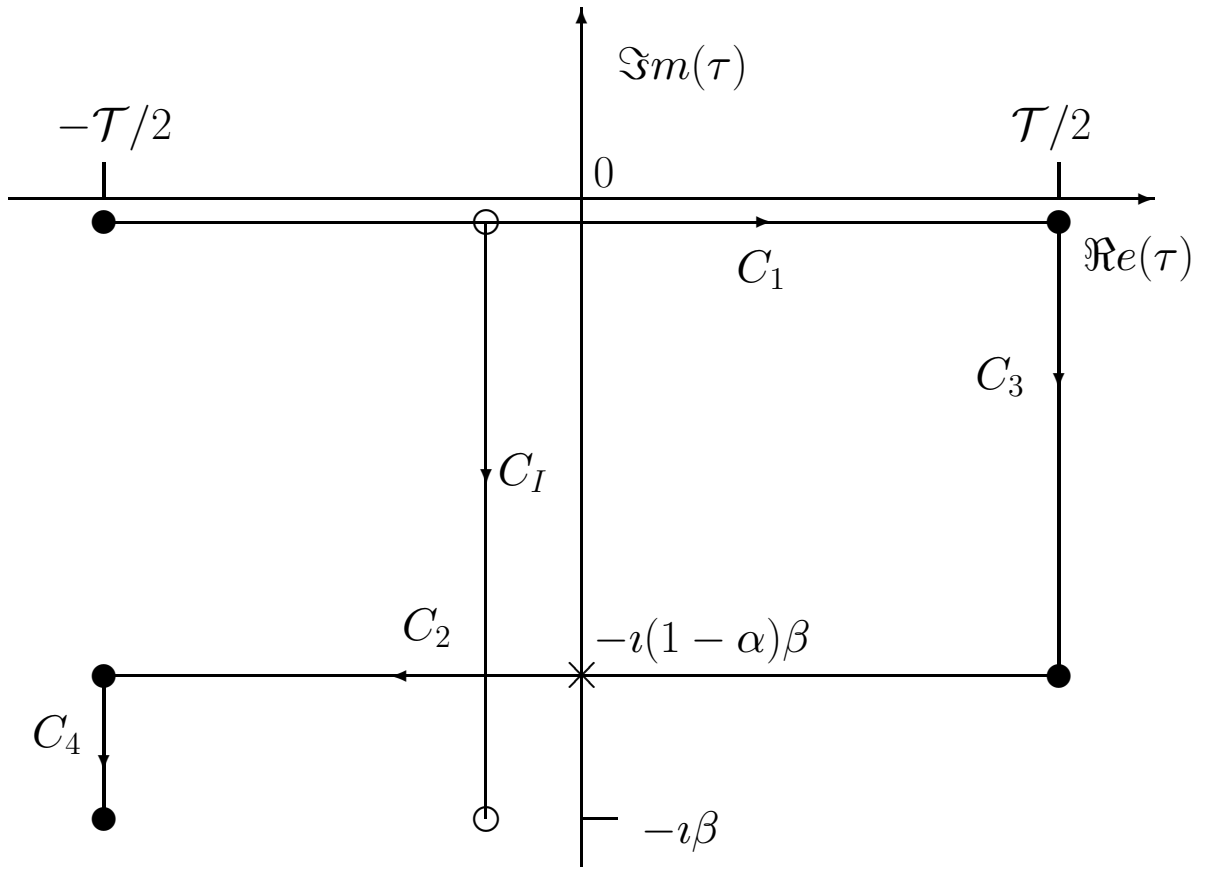

The ITF approach uses the curve $C=C_I$ of figure \ref{fitfortf}.
\typeout{LaTeX figure: ITFORTF - ITF and old RTF curves }
ITF is good for static quantities which include all the bulk macroscopic
information contained in the partition function.  It is harder to extract
dynamical information from ITF where real-times are
required\cite{TSErtgf} and thus difficult to extend the formalism to
non-equilibrium.  For these reasons a different curve $C$ was sought
which would lead to a real-time formalism, similar to that obtained in
the Thermo Field Dynamics approaches.

For a real-time formalism, one part
of $C$ must run along the whole of the real time axis.
The question is then how one completes
$C$ which must finish $-i\beta$ below its starting point.

The traditional curve for PORTF is $C=C_{old}=C_1 \oplus C_2 \oplus
C_3 \oplus C_4$ shown in figure \ref{fitfortf}.   The limit ${\cal T}
\rightarrow \infty$ is taken and the figure suggests that  $0
\leq\alpha\leq 1$.  Again no physical result depends on $\alpha$
suggesting any value can be taken provided we make use of the
periodic or anti-periodic boundary conditions when redrawing the curve.

The sections running parallel with the real time axis, $C_1,C_2$, are
not too bad to deal with as they can be parameterised by a
real time.  Thus a formalism close to familiar Minkowskii field theory
is obtained from these sections.   It is the vertical portions that are
unfamiliar and which cause problems.  In Thermo Field Dynamics
approaches there are only two real fields in their formalisms, and so
there is nothing which corresponds to the vertical sections $C_3,C_4$ of
$C_{old}$.  This suggests that only the $C_1$ and $C_2$ sections should be
kept in PORTF.  This is achieved in the literature through a process
called  factorisation, namely
\begin{equation}
Z[j] \stackrel{?}{\rightarrow} Z_{12}[j].Z_{34}[j]
\label{efactor}
\end{equation}
where in $Z_{ab}$ all the fields and
sources are limited to lie on ${C_{ab}}$,
\begin{eqnarray}
C_{ab} &=& C_a \oplus C_b  .
\end{eqnarray}
We can follow the usual derivation of this
result\cite{NS,LvW} in the path integral approach to PORTF.  In
this case
\begin{eqnarray}
Z[j] &=& exp\{ i \int_C d\tau \; V[ -i\frac{\partial}{\partial j} ] \}
\; \; . \; \; Z_0
\nonumber \\ {}
&=& exp\{ i \int_{C_{12}} d\tau \; V[ -i\frac{\partial}{\partial j} ] \}
exp\{ i \int_{C_{34}} d\tau \; V[ -i\frac{\partial}{\partial j} ] \}
\; \; . \; \; Z_0  ,
\label{eZ}
\\
Z_0[j] &=& exp\{ -\frac{i}{2} \int_c d\tau d{\tau}^\prime   \;
j(\tau) \Delta_c(\tau-{\tau}^\prime ) j({\tau}^\prime )   .
\label{eZ0}
\end{eqnarray}
Here the interaction terms in the Lagrangian are represented by the
functional $V$
and it is easy to see that the interaction part factorises in Eq. (\ref{eZ}).

The free
part is expressed in terms of the free propagator
$\Delta_C(\tau-{\tau}^\prime )$.  For many systems of interest the
propagator, $\Delta_C$,
tends to zero as the real part of the time difference tends to infinity
\begin{equation}
\lim_{ Re \{ \tau-{\tau}^\prime  \} \rightarrow \infty }
\Delta_c(  \tau-{\tau}^\prime )  = 0
\label{einfcorr}
\end{equation}

This is essential in the usual derivations of PORTF as the $C_3$ and
$C_4$ sections are at infinity whereas most of the $C_1$ and $C_2$ are
not.  For that perennial example, the relativistic scalar field, this
condition is satisfied provided the solution of the Klein-Gordon
equation is suitably regularised\cite{NS,LvW,Serot,PeEP2}.  The
Feynman $\epsilon$ regularisation achieves this.  Strictly, $\epsilon$
must be left finite till the end of the calculation.

However, $Z_0$ in Eq. (\ref{eZ0}) includes non-zero  contributions from
regions where one integral is near an end of $C_1$ or $C_2$ and the
other integral is running along $C_3$ or $C_4$.  In these situations
Eq. (\ref{einfcorr}) can not be used but such contributions to $Z_0$ must be
zero if factorisation is to be true.  The usual solution, termed an
``asymptotic condition'', is to say that the sources tend to zero for
times lying at the ends of $C_1$ and $C_2$\cite{NS,LvW}.
Unfortunately this turns out to be unacceptable.  The whole point of a
generating function and of the separation in Eq. (\ref{eZ}) is that the
sources $j$ must not be fixed.  In particular infinitesimal variations
are needed for the derivatives in the interaction terms in Eq. (\ref{eZ}).
It makes the expression Eq. (\ref{eZ}) meaningless
if $j$ is set to zero in Eq. (\ref{eZ0}) in
some regions.

One might try to get round this by switching off all interactions at the
ends of $C_1$ and $C_2$, a proper asymptotic condition c.f.\cite{Ab}.
However this means we have a time dependent Hamiltonian which
invalidates the fundamental assumption of an equilibrium situation.
It also means that the partition function, which is also a normalisation
factor for the connected Green functions, is being disturbed.
Finally, since the asymptotic condition is not used in
ITF, it seems strange that PORTF would need this extra boundary
condition.

In fact, however factorisation of Eq. (\ref{efactor}) is enforced,
it leads to a series of inconsistencies\cite{PeEP2}.
The inescapable conclusion is that the generating
function of PORTF  does {\em not} factorise.  However this does not
alter the fact that if we do drop the $C_3$ and $C_4$ parts of the old
PORTF curve, one obtains the same Feynman rules as one finds in the Thermo
Field Dynamics approaches are obtained\cite{NS,LvW}.

There are two ways out of this dilemma.  One is to keep working with
the whole of the old PORTF curve of figure \ref{fitfortf} and to just
use Eq. (\ref{einfcorr}) to get the usual answers.  For instance if this is
done problems encountered with the normalisation of Green
functions and with the partition function\cite{PeEP2} are
avoided.  However it is a bit cumbersome.

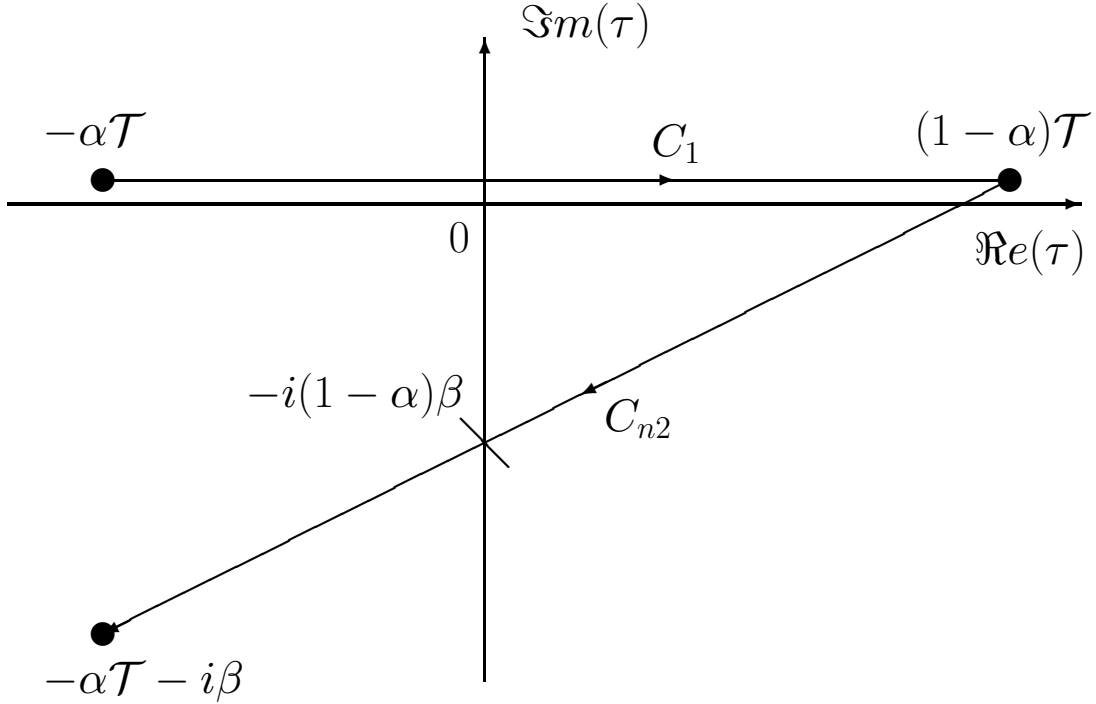
\begin{figure}[htb]
\setlength{\unitlength}{0.012500in}%
\begin{picture}(480,325)(100,315)
\thicklines
\put(160,340){\circle*{10}}
\put(160,530){\circle*{10}}
\put(540,530){\circle*{10}}
\put(320,320){\vector( 0, 1){270}}
\put(120,520){\vector( 1, 0){450}}
\put(160,530){\vector( 1, 0){240}}
\put(400,530){\line( 1, 0){140}}
\put(540,530){\vector(-2,-1){180}}
\put(360,440){\vector(-2,-1){200}}
\put(330,410){\line(-1, 1){ 20}}
\put(220,435){\makebox(0,0)[lb]{\Large $-i(1-\alpha)\beta$}}
\put(135,315){\makebox(0,0)[lb]{\Large $-\alpha {\cal T} - i \beta$}}
\put(135,545){\makebox(0,0)[lb]{\Large $-\alpha {\cal T}$}}
\put(500,545){\makebox(0,0)[lb]{\Large $(1-\alpha){\cal T}$}}
\put(305,500){\makebox(0,0)[lb]{\Large $0$}}
\put(525,495){\makebox(0,0)[lb]{\Large $\Re e (\tau)$}}
\put(335,590){\makebox(0,0)[lb]{\Large $\Im m (\tau)$}}
\put(390,540){\makebox(0,0)[lb]{\Large $C_{1}$}}
\put(370,425){\makebox(0,0)[lb]{\Large $C_{n2}$}}
\end{picture}
\caption{A new curve for real-time formalisms.}
\label{fnrtf}
\end{figure}

The alternative way out is to use a different curve for the PORTF.
Such a curve must run along the whole real axis to get all real times
accessible within the formalism.  There must be only two sections, each
parameterised by a real time parameter that runs between $\pm \infty$,
by analogy with Thermo Field Dynamics
methods and the existing successful Feynman rules for
real-time formalisms. Likewise there should be no dropping of any
sections.  One curve, $C_{new} = C_1 \oplus C_{n2}$ which satisfies these
criteria is where the curve is run straight back to the end point as
shown in figure \ref{fnrtf}.
\typeout{LaTeX figure FNRTF}
This is a special case of the curves presented elsewhere\cite{TSEnrtf}.
While we could of course
parameterise $C_{n2}$ in terms of a real parameter, the
gradient would surely hit you somewhere.
In this case one must remember that the ends
of the curve are going to be taken to infinity,
${\cal T}  \rightarrow \infty$,
so that the gradient is going to become zero.  The Feynman rules can be
derived in time coordinates, with ${\cal T} $ kept finite if required.  The
usual real-time Feynman rules in four-momentum space are obtained on
taking ${\cal T}  \rightarrow \infty$ and then doing the Fourier
transform\cite{TSEnrtf}.

It has been suggested that the vertical sections $C_3,C_4$ of the
traditional PORTF curve are essential to ensure the correct
energy-momentum Feynman rules\cite{Ni}.  That is they are supposed to
ensure that $n(\omega) \delta(k_0^2-\omega^2)$ terms in propagators ($n$
is the equilibrium number density) are replaced by the correct
$n(|k_0|) \delta(k_0^2-\omega^2)$ terms.  However the $n(\omega)$ form
is simply incorrect because then the propagator does not satisfy the
boundary conditions of equilibrium Green
functions\cite{LvW}\tnote{LvW\cite{LvW}, pp172}.   It
is therefore not a problem that the new curve does not have these
vertical pieces present in the old curve, the problem is one of how to
correctly  take the Fourier transform of a free propagator.

The gradient of $C_{n2}$ of the new curve can be ignored in
calculating thermal Green
functions provided Eq. (\ref{einfcorr}) holds\cite{PeEP2}.  It can
not, however, be ignored when calculating vacuum diagrams e.g. in a
diagrammatic expansion of the partition function or Free
energy\cite{PeEP2}.  In this case one can use the trick\cite{TSEz}
in which a vacuum diagram is treated as a time integral
multiplied by a tadpole type Green function diagram.

It turns out that the $C_3,C_4$ vertical sections of the old PORTF
curve can be ignored in exactly the same circumstances that the
gradient of the new curve can be neglected.  Likewise when the
gradient must be included, the effect of the vertical pieces is
important.   Thus the same problems are hiding in both methods.   The
new curve does however allow a much simpler and cleaner derivation of
the Feynman rules.  It also emphasises that there is no need for any
sort of ``asymptotic condition''.  This is true whatever sort of
curve is used in path-ordered methods despite what is found in the
literature on PORTF using the old curve. In particular parts of the
standard PORTF derivations are clearly wrong yet the answers the
resulting formalism gives is correct.

Finally when the vertical sections are needed,  it not easy to see how
to include their effects in the old PORTF whereas it is straightforward
to keep the gradient terms in the new approach when they are needed.
This is most important when long time correlations are not zero, i.e.
Eq. (\ref{einfcorr}) no longer holds for some of the fields in the problem.
This occurs in certain models such as the Anderson
model,\tnote{LvW\cite{LvW} pp.173}\cite{LvW}, and in certain physical
situations e.g. near critical points.  The new approach to PORTF using
the curve of figure \ref{fnrtf} is therefore likely to be of practical
benefit.

I would like to thank the Royal Society for their support through a
University Research Fellowship.  I am happy to acknowledge useful
conversations with many of those attending the conference, especially in
this context, A. Pearson and C. van Weert.  I would finally like to
thank the organisers for putting together such a useful and enjoyable
meeting.

\typeout{--- references ---}

\end{document}